\documentstyle[prb,aps,tighten,preprint,eqsecnum]{revtex} 
\begin{document}
\title{Quantum mechanics with coordinate-dependent mass}

\author{A.~V.~Kolesnikov$^\dagger$ and A.~P.~Silin$^\ddagger$}

\address{$^\dagger$ Fakult\"{a}t f\"{u}r Physik und Astronomie, \\
Ruhr-Universit\"{a}t Bochum,\\Universit\"{a}tsstr. 150, Bochum, Germany}

\address{$^\ddagger$ Tamm Theoretical Department of the \\ Lebedev Physical
Institute,~RAS, \\  Leninskii pr. 53, 117924, Moscow,  Russia}
\maketitle 

\begin{abstract}
We study a motion of quantum particles, whose properties depend on 
one coordinate so that they can move freely in the perpendicular direction.
A rotationally-symmetric Hamiltonian is derived  and applied 
to study a general interface formed between two semiconductors. We predict a 
new type of electron states, localized at the interface. They appear whenever 
the two bulk dispersions intersect. These shallow states lie near the point of 
intersection and are restricted to a finite range of perpendicular momentum. 
The scattering of carriers by the interface  is discussed.
\end{abstract}  

\section{Introduction} 
                                           
Highly developed methods of crystal growth allow building of mesoscopic
systems with spatially varying parameters. The motion of carriers in such a 
system is described within the effective mass theory (EMT) by a quantum 
equation. Depending on the semiconductors chosen, this can be either 
Schr\"odinger-like for wide-gap materials, or Dirac-like or more complicated 
for narrow-gap semiconductors \cite{SSPBa}. Thus, there is a way to realize 
in practice the quantum mechanics with parameters depending on coordinates. 
All information about the periodic quickly varying potential of the lattice
is contained in the effective mass or the ``velocity of light'' -- the  
interband matrix element. 

The foundation of the phenomenological EMT is based on the envelope function 
approximation (EFA). The total wave function is represented as the product of 
the Bloch function and a slowly varying envelope function; it is the latter 
which enters the quantum equation. The commonly used heuristic EFA \cite{Babuch} 
exploits the fact that the interband matrix element does not change much
in many important for practice III-V semiconductors. This leads to a reasonable
assumption that the Bloch function can be chosen to be identical through the
whole system. The resulting EMT is equivalent to the conventional quantum 
mechanics, e.g. the envelope function should be continuous at any heterojunction.
However, the interband matrix element does vary for different semiconductors
\cite{Manual}. Recently, a new exact version of the EFA was suggested \cite{B},
where only the total wave function has to be continuous and the envelope 
function does not need necessarily to fulfill this requirement.

In Ref. \cite{Tich}, first evidence indicated the existence of new interface
states, impossible in the conventional quantum mechanics. The further 
investigation \cite{WE1} revealed a novel effect for the coordinate-dependent 
two-band Dirac Hamiltonian: A localization can occur at a step-like potential 
(junction), provided that the free motion along the junction is accounted for.
The envelope wave function was shown to be discontinuous for these states.
In the present paper, we study an arbitrary junction of two wide-gap 
semiconductors. Also here, the free motion 
along the junction leads to binding in its perpendicular direction.

Although graded crystals with very different dependences of the effective
mass $m(z)$ on the coordinate $z$ can be grown, the most attention has been 
drawn to abrupt heterostructures. This is probably because one does not need 
to use a Hamiltonian with coordinate-dependent kinetic term (see below). 
Instead, a problem of choosing the 
matching conditions (MC), connecting the wave functions on either side of a 
junction, is met. Usually they are derived from the requirements of the 
continuity of the probability and its current at the junction. These 
requirements, although being correct, are not constructive. Indeed, they are 
meaningless for matching of exponentially growing or decaying solutions as 
their current vanishes. On the other hand, incident and scattered waves appear 
in the propagating case. Two continuity equations are not enough to fix the 
MC and two constants of the scattering problem. 

An alternative approach of deriving the MC directly from the Hamiltonian is 
suggested in Ref. \cite{WE1}: The abrupt junction can be treated as a limiting 
case of a smooth one, when its ``smoothness'' length tends to zero. For the 
graded case, the Hermitian Dirac operator is derived and from it the MC are 
extracted. Apart of appealing to intuition, such a method is shown to be 
equivalent to the conventional scheme. The Dirac Hamiltonian is 
convenient to handle with, because it contains first-order derivatives, so 
that it can be easily symmetrized and it is originally rotationally-symmetric. In 
the present paper, we follow the above-described approach and develop an 
effective way to derive Hermitian, rotationally-invariant one-band Schr\"odinger 
Hamiltonian. The MC depend on the free motion, since it is incorporated in the 
Hamiltonian.

The paper is organized as follows: In Sec. II, we derive the Hamiltonian and
obtain from it the matching conditions. In Sec. III, the dispersion of the
interface states localized at the heterojunction is found and the scattering 
of the carriers by the junction is discussed. Sec. IV summarizes the results.

\section{Rotationally-symmetric Schr\"odinger Hamiltonian}

We study a junction of two semiconductors with the effective masses $m_i$ 
and the bandgaps $2\Delta_i$, dependent on one coordinate $z$, $i=1$ for $z<0$
and $i=2$ for $z>0$ (see Fig. 1). Since all parameters depend on $z$ only, 
the free motion in the perpendicular direction $x$ has a good quantum 
number $k_\bot$ so that the wave function is $\psi \sim \exp(i k_\bot x)$. 
(Hereinafter we put $\hbar=1$.)
Let us start with the case of vanishing free motion $k_\bot=0$; we
consider $k_\bot \neq 0$ later. The problem one immediately meets while 
studying a graded  crystal with some function $m(z)$ is that the 
kinetic term $T=\partial^2_z/2m$ is not Hermitian. Its most general form, 
after symmetrization, was found \cite{M-B}  to equal 
\begin{eqnarray} T=m^\alpha\partial_z m^\beta\partial_z m^\alpha/2=
\frac{1}{2m} \left[\partial^2_z-\frac{m^\prime}{m}\partial_z +\alpha \beta
\left(\frac{m^\prime}{m}\right)^2 +\alpha \frac{m^{\prime \prime}}{m}\right]
\enspace , \label{a1} \end{eqnarray}
with constants $\alpha$ and $\beta$ such that $2\alpha+\beta=-1$.
Note the appearance of singular terms for abrupt junctions $m^{\prime
\prime}\simeq\delta^\prime(z)\Delta_m$ and $(m^\prime)^2\simeq \delta^2(z)(
\Delta_m)^2$, where $\delta(z)$ is the Dirac delta-function and $\Delta_m=
m_2-m_1$. These singularities lead to a discontinuity of the wave function. 
Representing the wave function $\psi(z)=\varphi(z) f(z)$ 
as a product of the continuous part $f(+0)=f(-0)$ and the function 
$\varphi(z)$, which jumps at the junction, $\Delta_{\varphi}=\varphi(+0)
-\varphi(-0)$, one obtains %selects the singular contributions 
$\psi^\prime \simeq f\Delta_{\varphi}\delta(z)$ %+\varphi f^\prime$ 
and $\psi^{\prime\prime} \simeq f\Delta_{\varphi}\delta^\prime(z)+2
\Delta_{\varphi} \delta(z)f^\prime$. %+\varphi f^{\prime\prime}$. 
Substituting these expressions into Eq.~(\ref{a1}) and selecting the singular 
contributions, proportional to $\delta^\prime(z)$ and $\delta^2(z)$, one 
finds that all the singularities in Eq. (\ref{a1}) are compensated under the 
following conditions: $\Delta_{\varphi}/\varphi+\alpha\Delta_m/m=0$ and  
$ \alpha\beta(\Delta_m)/m-\Delta_{\varphi}/\varphi=0$. These give
either $\beta=-1$ or $\alpha=0$. Recalling that  $2\alpha+\beta=-1$, one 
notices that both possibilities lead to the same effective Hamiltonian 
$T(z)=\partial_z (1/2m) \partial_z$, acting on the continuous function $f(z)$.
This form was already derived in many works \cite{1,2,3,4,5}, where arguments
different from ours were used.

The next step is to incorporate the free motion along the surface. Up to our 
knowledge, the Hermitian rotationally-symmetric form of the Schr\"odinger 
Hamiltonian has not been written yet. The only form, being 
simplified for $k_\bot=0$ to the one-dimensional case, recovering the Laplace 
operator for constant $m$, and invariant with respect to orthogonal rotations 
involving $z$ and $x$ is given by \begin{eqnarray}T=(\partial_z-i\lambda
\partial_x) \frac{1}{2m} (\partial_z+i\lambda\partial_x)\enspace , \label{a2}
\end{eqnarray} where $\lambda=\pm 1$. The physical meaning of the value 
$\lambda$ will be elucidated later, from the comparison of the one-band and 
the two-band Hamiltonians. The appearance of $\lambda$ is due to the fact that 
the axial symmetry possessed  by the heterojunction is broken as soon as $k_\bot 
\neq 0$. Two signs of $\lambda$ formally 
account for two possible ways of writing the kinetic form symmetrically.
Acting by the operator $T$ on the function $f=g(z)\exp(ik_\bot x)$, we derive 
the eigenvalue equation for the energy $\epsilon$ and half the energy gap, 
$\Delta(z)$, playing the role of the potential at $k_\bot=0$: 
\begin{eqnarray}\left (\frac{1}{2m}\partial^2_z-\frac{m^\prime}{2m^2} \partial_z
+\lambda \frac{m^\prime}{2m^2}k_\bot-\frac{k^2_\bot}{2m}-\Delta+\epsilon \right 
)g=0\enspace . \label{a3} \end{eqnarray} The presence of the third term in 
Eq.~(\ref{a3}), originating from the mixed derivative $\partial_z(1/m) \partial_x$ 
in Eq.~(\ref{a2}), will be crucial for the following discussion. Let us search for 
solutions of Eq.~(\ref{a3}) in the form $g=\exp[\int{\rm d}z \,\kappa(z)]$. Far 
from the junction, $z\rightarrow\pm \infty$, the second and the third terms  of 
Eq.~(\ref{a3}) vanish. The remaining terms determine the values $\kappa_i=\pm
\sqrt{2m_i(\Delta_i-\epsilon)+k^2_\bot}$, nothing else but the (imaginary) wave 
numbers of the two bulk semiconductors. The terms  $\kappa^\prime/2m-\kappa 
m^\prime/2m^2+\lambda k_\bot m^\prime/2m^2$ prevail at the junction. To avoid 
unphysical discontinuities in the abrupt case, we require $[(\kappa-\lambda 
k_\bot)/m]^\prime=0$, recovering thus the matching conditions \begin{eqnarray}
\left[(\kappa-\lambda k_\bot)/m\right ]_1=\left [(-\kappa-\lambda k_\bot)/m
\right]_2 \enspace . \label{a4} \end{eqnarray} The sign of $\kappa_1$ and 
$\kappa_2$ is chosen in the way to ensure a localized wave function. For the 
case $m=const$, these matching conditions coincide 
with those of Bastard \cite{BMC} and do not depend on $k_\bot$.

\section{Interface states}

If the energy $\epsilon$ is measured from $(\Delta_1+\Delta_2)/2$, one 
rewrites $\kappa^2_i=k^2_\bot-2m_i(\epsilon\pm\Delta)$, where the sign ``$+$'' 
corresponds to $i=1$, ``$-$'' to $i=2$ and $\Delta=(\Delta_1-
\Delta_2)/2$. Equation (\ref{a4}), together with the bulk values of $\kappa_i$, 
yields  \begin{eqnarray} \epsilon(k_\bot)=-P(m_1+m_2)+\lambda 2 k_\bot \sqrt{P}
\enspace , \label{b1}  \end{eqnarray} where $P=|\Delta/ (m_2-m_1)|$. To obtain 
Eq.~(\ref{b1}) one squares Eq.~(\ref{a4}), therefore it should be found which 
branch of Eq.~(\ref{b1}) in which region represents the solution of Eq.~(\ref{a4}) 
and corresponds to localized interface states. A simple analysis reveals that the 
states are localized in the region $k^{\rm min}_{\bot}<k_\bot<k^{\rm max}_\bot$, 
between minimal and maximal values of $k^2_{i\,\bot}=4 m^2_i P$. At $k_{i\,\bot}$, 
the curve $\epsilon(k_\bot)$ is tangential to the bulk dispersions $\epsilon_{i}=
\mp\Delta+k^2_\bot/2m_i$. The necessary condition for the interface  states to 
exist is $(m_1-m_2)(\Delta_1-\Delta_2)>0$, i.e. the bulk dispersions must 
intersect. The dispersion $\epsilon(k_\bot)$, Eq.~(\ref{b1}), and the bulk 
dispersions $\epsilon_i$ are shown in  Fig. 2  for the following parameters: 
$\Delta_1=1.2$ eV, $\Delta_2=1.4$ eV, $m_1=0.01\; m_0$, $m_2=0.02\; m_0$ ($m_0$ is 
the free mass of the electron).

It follows from Eq.~(\ref{a4}) that $\lambda k_\bot(m_2-m_1)$ is positive. Recalling 
that the signs of $\lambda$ ``$+$'' and ``$-$'' originate from two possible ways 
of writing the Hamiltonian (\ref{a2}), we conclude that the opposite signs in 
Eq.~(\ref{a4}) correspond to localized states for positive and negative $k_\bot$. 
That is, if $m_2>m_1$, then $k_\bot>0$ is localized for $\lambda=+1$ in Eq.~(\ref{a2}) 
and $k_\bot<0$ is localized for $\lambda=-1$. The signs in Eq.~(\ref{b1}) have to be 
chosen accordingly.

Since both values of $\lambda=\pm1$ are equivalent, the degeneracy of bulk dispersion 
of the Schr\"odinger Hamiltonian with respect to the sign of $k_\bot$ is preserved 
for the localized states: $\epsilon(k_\bot)=\epsilon(-k_\bot)$. 
This differs for the two-band model with the non-relativistic Dirac Hamiltonian 
%\cite{WE1,WE2}  \begin{eqnarray} \left ( 
%\begin{array}{cc}\Delta - \epsilon_{_{\lambda}} 
%& v \lambda k_\bot - v
%\partial_z \\ v \lambda k_\bot + v \partial_z & -\Delta -
%\epsilon_{_{\lambda}} \end{array} \right ) \left 
%( \begin{array}{c} 
%\psi_{_\lambda}\\ \chi_{_\lambda}\end{array} \right ) 
%= 0\enspace ,\end{eqnarray}  
%where $\psi_{_\lambda}$ and $\chi_{_\lambda}$ are two 
%wave functions of the two 
%hybridized bands, $v$ is the interband velocity matrix 
%element, and  $\Delta$ is 
%half the energy gap. Here, 
\cite{WE2}, where the eigenvalues of energy $\epsilon$ are classified 
(apart of momentum $k_\bot$) by eigenvalues of helicity $\lambda=\pm1$. Here, the 
initial degeneracy of the bulk dispersion is lifted for the interface states: 
$\epsilon_\lambda(k_\bot)=\epsilon_{-\lambda}(-k_\bot)$ \cite{WE1}. 
%Eliminating one of the wave functions $\psi_{_\lambda}$ or $\chi_{_\lambda}$, one 
%arrives at the Hamiltonian with 
It is not difficult to check that the Klein-Gordon equation, obtained in the 
two-band model, contains the kinetic term analogous to Eq.~(\ref{a2}). Thus, the 
value $\lambda$ in Eq.~(\ref{a2}) corresponds to helicity in the two-band model. In 
discussing further similarities of the one- and two-band description, it is worth 
noticing following: Although qualitatively the dispersion of the interface states is 
similar in both models, corresponding to localized states between tangency points 
with the two bulk dispersions, the dispersion in the two-band model is nonlinear 
due to different bulk spectrum.

%\section{Scattering by the junction}

Let us now briefly discuss the scattering of carriers by the interface. Like for
the bound states, it is convenient to represent the wave function in the form 
$g(z)=\exp[\pm i \int {\rm d} z \,\kappa(z)]$. Then for $z<0$ it will be the sum
of incident and reflected waves $g=\exp[i\int {\rm d} z \, \kappa_1(z)]+B
\exp[-i\int {\rm d} z \, \kappa_1(z)]$ and for $z>0$ the transmitted wave is
$g=A\exp[i\int {\rm d} z \, \kappa_2(z)]$, where $\kappa^2_i=2m_i(\epsilon\pm
\Delta)-k^2_\bot$. The condition of continuity of the wave function $g$ yields 
$1+B=A$ and the analog of Eq.~(\ref{a4}) reads $[(k_\bot+i\kappa)/m+B(k_\bot-i\kappa)/
m]|_1=A[(k_\bot+i\kappa)/m]|_2$. The transmission coefficient $D(\epsilon)=|A|^2
\kappa_2/\kappa_1$ is given by \begin{eqnarray} D(\epsilon)=4\kappa_1\kappa_2m_2^2/
[k^2_\bot(m_1-m_2)^2+(\kappa_1m_2+\kappa_2m_1)^2] \enspace . \label{b2} \end{eqnarray}

Far away from the point of intersection, $k^2_\bot \ll4m_1m_2P$ or $k^2_\bot 
\gg 4m_1m_2P$, function $D(\epsilon)$ vanishes as square root
as the energy tends to its threshold value (one of the bulk dispersions): $D(
\epsilon) \sim \kappa_i$. On the other hand, near the point of intersection, 
$k^2_\bot \approx 4m_1m_2P$, the function $D(\epsilon)$ vanishes linearly: 
$D(\epsilon) \sim \epsilon - (m_1+m_2)P$. This behaviour is typical for the 
conventional quantum mechanics: the coefficient of penetration $D(\epsilon)$ is 
governed by the asymptotic behavior of the potential $U(z)$ only. If the value 
$U(+\infty)-U(-\infty)$ is zero than $D(\epsilon)$ is linear at the penetration 
threshold. In the opposite case it vanishes as square root \cite{L-L}. 

\section{Conclusions}

It is worth emphasizing that it is the free motion along the junction, 
%namely its mixing with the limited motion in  $z$-direction, 
which leads to the effects discussed. 
Although for $k_\bot=0$ we have a step-like potential $U(z)=\Delta_i$, its role 
for $k_\bot \neq 0$ is taken over by some more complicated function 
$U^\pm(z, k_\bot)$. Substituting into Eq. (\ref{a3}) $g(z)=m^{1/2}y(z)$,
we obtain a Schr\"odinger equation for the function $y(z)$ with the potential
\begin{eqnarray} U^\pm(z, k_\bot)=3m^{\prime\,2}/8m^3-m^{\prime \prime}/
4m^2-\lambda k_\bot m^\prime/2m^2+k^2_\bot/2m+\Delta \enspace . \label{c1} 
\end{eqnarray} The potential $U^\pm$ for $m(z)=(m_1+m_2)/2+(m_2-m_1)\tan(z/l)/2$ 
with the smoothness parameter $l$ is presented in Fig. 3 as a function of $z$ for 
several values of $k_\bot$ and $l$. The asymptotic behavior of the potential is 
preserved for any smooth $m(z)$, $U(\pm\infty)=\Delta_i+k^2_\bot/2m_i$. If the 
bulk dispersions intersect, then at $k_\bot$, corresponding to the point of 
intersection, we have no step in the potential any more: $U(+\infty)=U(-\infty)$. 
As it is known from quantum mechanics, any potential with this property and 
containing a well, has at least one localized level. This level becomes shallow, 
if the potential changes rapidly, $l \rightarrow 0$, (e.g. a $\delta$-function). 
For the abrupt potential, the part of the well is played by the matching 
conditions Eq.~(\ref{a4}). For $k_\bot \neq 0$ and small $\kappa_i$, i.e. near the 
intersection point of the bulk dispersions, Eq.~(\ref{a4}) has a solution, describing 
a {\it shallow} level. 
%This is impossible for the conventional quantum mechanics, $m=const$. 
The scattering by the junction is interpreted analogously.

The states considered lie above the band edge, being embedded in the continuum 
spectrum. However, they are real localized states, not resonant ones, since they
lie in the energy gap of the whole system: Their energies are below the bulk 
dispersions, corresponding to the same value of $k_\bot$. Naturally, many-particle 
effects, due to e.g. impurities and the boundary roughness could strongly affect 
these states. However, qualitatively it is clear that as long as these can be 
treated as elastic scatterer, the states preserve because in our analysis the 
absolute value of $k_\bot$ was essential, but not its direction.

%This implies a change of sign of the logarithmic derivative. 

In summary, we have derived a rotationally-invariant effective one-band 
Hamiltonian, which
mass depends on one coordinate. We applied it to derive the matching  conditions 
for a single general junction of two semiconductors. They depend on the free 
motion along the junction, due to its mixing with the motion in the 
direction of growth. If the bulk dispersions of the two semiconductors intersect, 
then shallow localized states occur. The problem studied can be classified as an 
example of weak localization. The essential condition for the existence of the 
states, the intersection of the bulk dispersion curves, holds for a large variety   
of semiconductors \cite{Manual}. Although more investigation is needed to 
understand in detail the many-particle effects, we believe that the states 
%discussed can be observed experimentally.
are not crucially sensitive to them.

\section{Acknowledgements}

We are indebted to K.~B.~Efetov and S.~G.~Tikhodeev for useful discussions. APS 
was supported in part by the Russian Foundation for the Fundamental Research 
under projects No. 96-02-16701, 97-02-16346, by Russian Science Ministry under 
project No. 97-1087, INTAS 96-0398. AVK acknowledges a support from the 
Sonderforschungsbereich 237 ``Unordnung und grosse Fluktuationen''.

%\newpage
 
%\end{twocolumn} 

%\begin{references}

\bigskip
%\begin{figure}\centerline
\newpage  %\noindent
\centerline{\large  FIGURE CAPTIONS}

\bigskip
\noindent
Figure 1.
Band-edge profile of an heterojunction. The energy gap is shaded. Because of 
symmetry, only the region of energy near the conduction bands is studied.

\smallskip
\noindent
Figure 2.
Dispersion $\epsilon(k_\bot)$, Eq.~(\ref{b1}), of the system shown in Fig. 1 
(solid curve; for parameters, see text). Dotted curves: bulk  dispersions 
$\epsilon_{1,\; 2} (k_\bot)$. Energy is measured from the middle of the bandgap.
States are localized between tangency points $k_{i\,\bot}$ with the two bulk 
dispersions (heavy solid line). It is understood that similar curves occurs for 
$k_\bot<0$ and near the valence band.

\smallskip
\noindent
Figure 3.
Effective potential $U^+$, Eq.~(\ref{c1}), of the smooth junction; for parameters:
1) $k_\bot=10^6$ cm$^{-1}$, $l=1.5\times10^5$ cm$^{-1}$; 2) $k_\bot=5\times 10^6$ 
cm$^{-1}$, $l=1.5\times 10^5$ cm$^{-1}$; 3) $k_\bot=5\times10^6$ cm$^{-1}$, $l= 
10^5$ cm$^{-1}$; 4) $k_\bot=6.2\times 10^6$ cm$^{-1}$, $l=1.5\times10^5$ cm$^{-1}$.  
Note appearance of a barrier for $\lambda=-1$, $U^-$ (curve 5); parameters are 
the same as for curve 2.

\newpage
\begin{center}
  \unitlength1cm
  \begin{minipage}[t]{12cm}
%\vspace*{-2cm}
  \begin{picture}(0,5)
     \put(-4.,-21.5)
    {\includegraphics{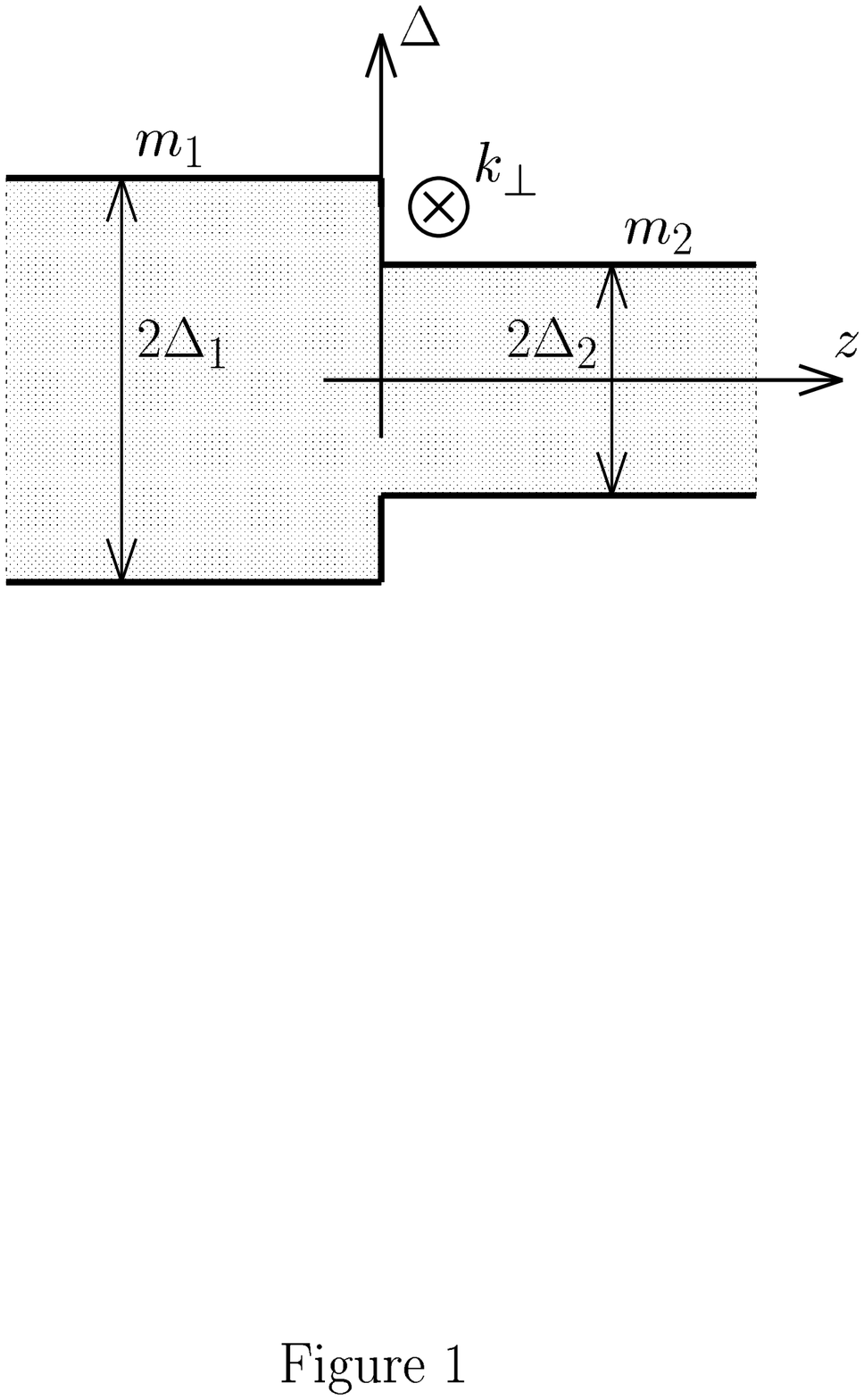}}
 \end{picture}
  \end{minipage}
\end{center}            

\newpage
\begin{center}
  \unitlength1cm
  \begin{minipage}[t]{12cm}
%\vspace*{-2cm}
  \begin{picture}(0,5)
     \put(-4.,-21.)
    {\includegraphics{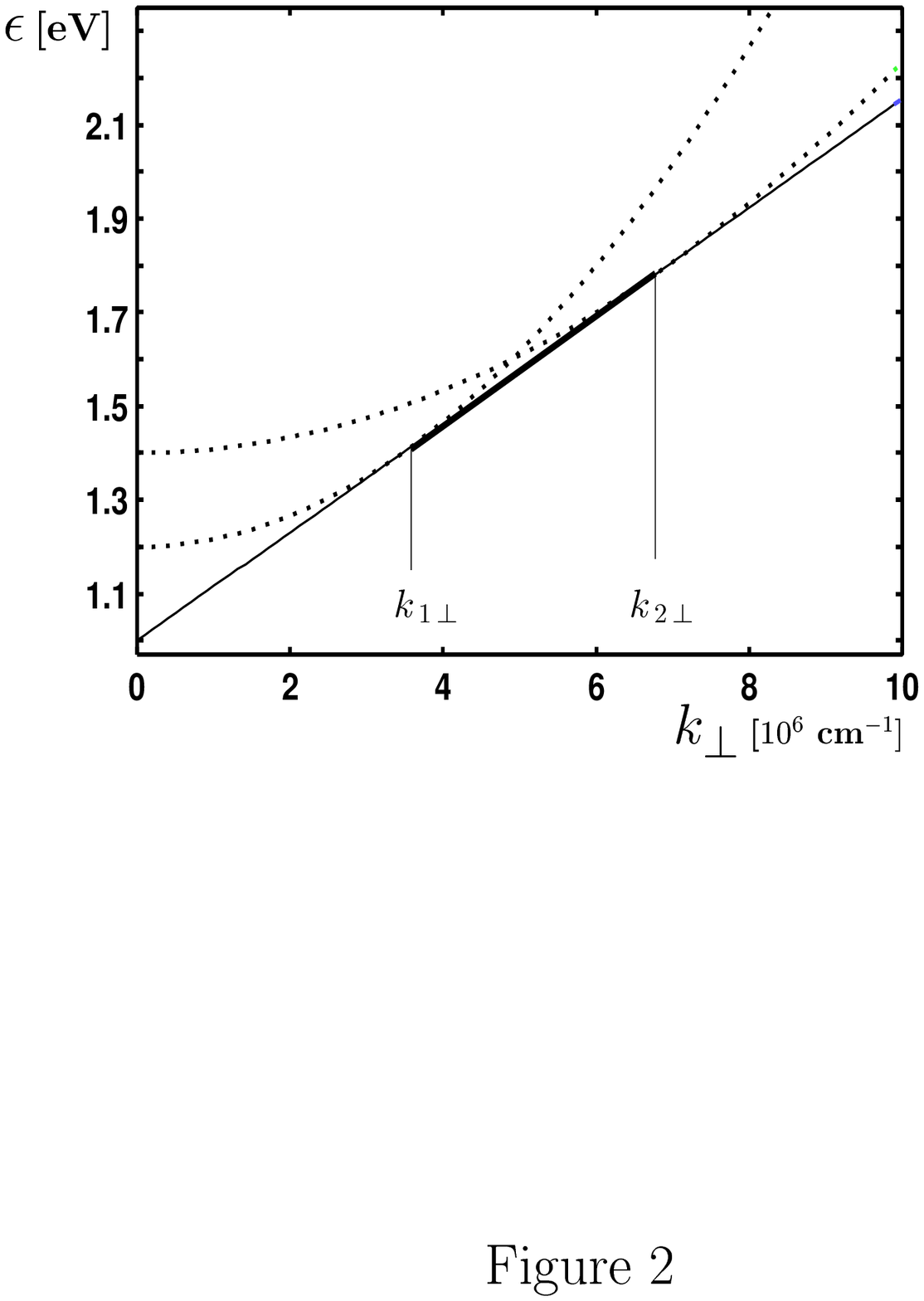}}
 \end{picture}
  \end{minipage}
\end{center}            

\newpage
\begin{center}
  \unitlength1cm
  \begin{minipage}[t]{12cm}
%\vspace*{-2cm}
  \begin{picture}(0,5)
     \put(-4.,-21.)
    {\includegraphics{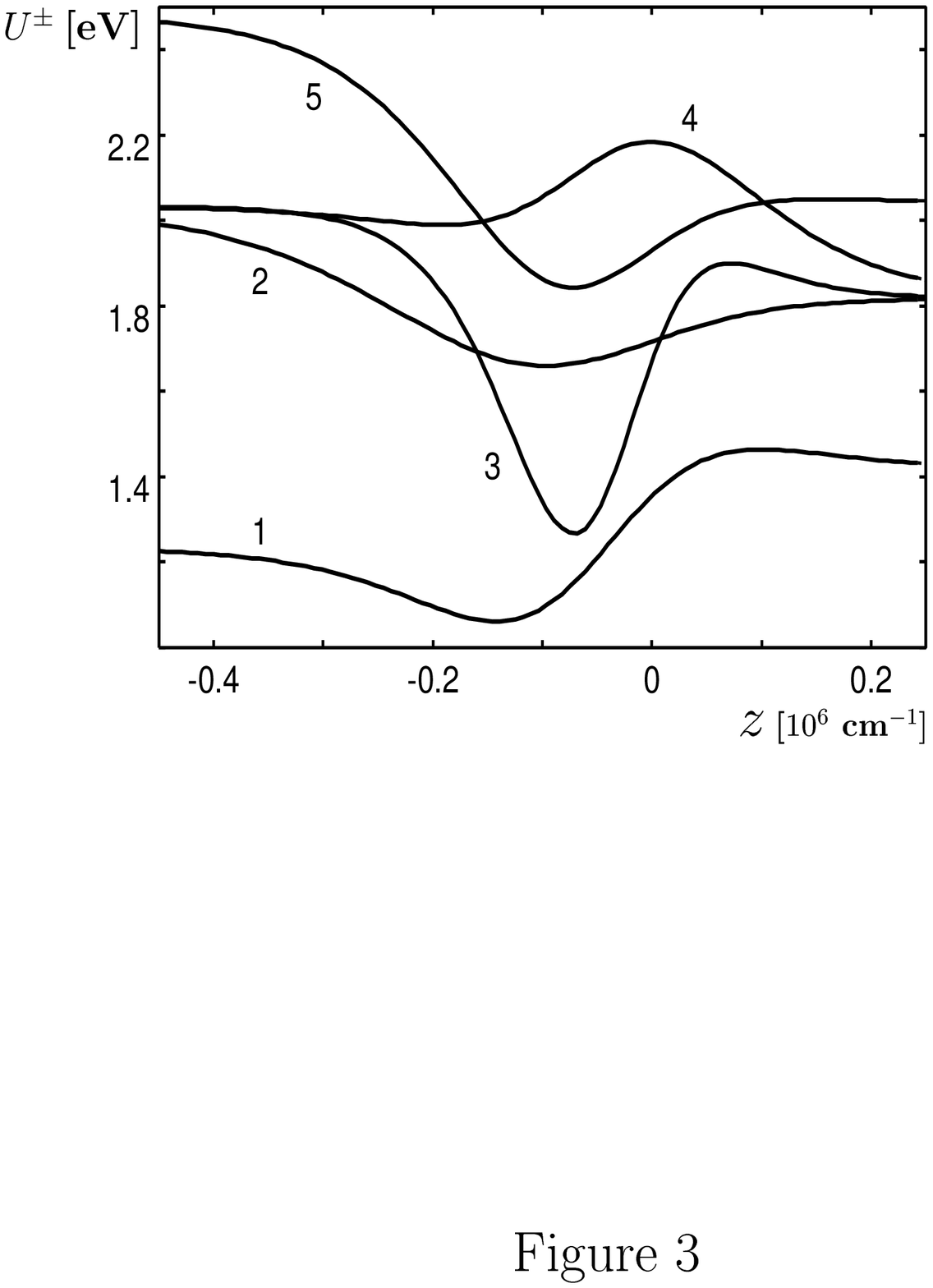}}
 \end{picture}
  \end{minipage}
\end{center}

\end{document}